\documentclass[sigconf]{acmart}

\usepackage{bigstrut}
\usepackage{booktabs}
\usepackage{multirow}
\usepackage{microtype}
\usepackage{graphicx}
\usepackage{subfig}
\usepackage{acmart-taps}
\usepackage{balance}
\usepackage{amsthm}
\usepackage{ifsym}
\usepackage{array}
\usepackage{clrscode}
\usepackage{multicol}
\usepackage{float}
\usepackage{newfloat}
\usepackage{color}
\usepackage{xcolor}
\usepackage{amsopn}
\usepackage{mathrsfs}
\usepackage{mathtools}
\usepackage{amsmath}
\usepackage{arydshln}
\usepackage{blkarray}
\usepackage{enumerate}
\usepackage{rotating}
\usepackage{bm}
\usepackage{wrapfig}
\usepackage{ragged2e}
\usepackage{makecell}
\usepackage{enumitem}
\usepackage{url}
\usepackage{xspace}
\usepackage{comment}
\usepackage{changepage}
\usepackage{algorithm}
\usepackage{algpseudocode} 
\usepackage{algorithmicx} 
\usepackage[ruled, vlined, nofillcomment, linesnumbered, algo2e]{algorithm2e}

\definecolor{codegreen}{rgb}{0,0.3,0.6}
\definecolor{codegray}{rgb}{0.5,0.5,0.5}

\newcommand{\ie}{\emph{i.e.,}\xspace}
\newcommand{\eg}{\emph{e.g.,}\xspace}

\newcommand{\paratitle}[1]{\vspace{1.5ex}\noindent\textbf{#1}}
\newcommand{\wrt}{w.r.t.\xspace}

\newcommand{\ignore}[1]{}

\AtBeginDocument{%
  \providecommand\BibTeX{{%
    \normalfont B\kern-0.5em{\scshape i\kern-0.25em b}\kern-0.8em\TeX}}}

\setcopyright{acmlicensed}
\copyrightyear{2025}
\acmYear{2025}
\setcopyright{acmlicensed}\acmConference[RecSys '25]{Proceedings of the Nineteenth ACM Conference on Recommender Systems}{September 22--26, 2025}{Prague, Czech Republic}
\acmBooktitle{Proceedings of the Nineteenth ACM Conference on Recommender Systems (RecSys '25), September 22--26, 2025, Prague, Czech Republic}
\acmDOI{10.1145/3705328.3748075}
\acmISBN{979-8-4007-1364-4/2025/09}

\begin{document}

\title[Enhancing Sequential Recommender with Large Language Models\\ for Joint Video and Comment Recommendation]{Enhancing Sequential Recommender with Large Language Models for Joint Video and Comment Recommendation}

\author{Bowen Zheng}
\orcid{0009-0002-3010-7899}
\email{bwzheng0324@ruc.edu.cn}
\affiliation{%
    \institution{Renmin University of China}
    \city{Beijing}
    \country{China}
}

\author{Zihan Lin}
\orcid{0000-0002-6877-4470}
\email{linzihan03@kuaishou.com}
\affiliation{%
    \institution{Kuaishou Technology Co., Ltd.}
    \city{Beijing}
    \country{China}
}

\author{Enze Liu}
\orcid{0009-0007-8344-4780}
\email{enzeeleo@gmail.com}
\affiliation{%
    \institution{Renmin University of China}
    \city{Beijing}
    \country{China}
}

\author{Chen Yang}
\orcid{0000-0001-5228-3426}
\email{flust@ruc.edu.cn}
\affiliation{%
    \institution{Renmin University of China}
    \city{Beijing}
    \country{China}
}

\author{Enyang Bai}
\email{baienyang@kuaishou.com}
\affiliation{%
    \institution{Kuaishou Technology Co., Ltd.}
    \city{Beijing}
    \country{China}
}

\author{Cheng Ling}
\email{lingcheng@kuaishou.com}
\affiliation{%
    \institution{Kuaishou Technology Co., Ltd.}
    \city{Beijing}
    \country{China}
}

\author{Han Li}
\email{lihan08@kuaishou.com}
\affiliation{%
    \institution{Kuaishou Technology Co., Ltd.}
    \city{Beijing}
    \country{China}
}

\author{Wayne Xin Zhao$^\dagger$}
\orcid{0000-0002-8333-6196}
\email{batmanfly@gmail.com}
\affiliation{
    \institution{Renmin University of China}
    \city{Beijing}
    \country{China}
}

\author{Ji-Rong Wen}
\orcid{0000-0002-9777-9676}
\email{jrwen@ruc.edu.cn}
\affiliation{
    \institution{Renmin University of China}
    \city{Beijing}
    \country{China}
}

\thanks{$\dagger$ \ Corresponding author.}

\renewcommand{\shortauthors}{Bowen Zheng, et al.}

\begin{abstract}
Nowadays, reading or writing comments on captivating videos has emerged as a critical part of the viewing experience on online video platforms.
However, existing recommender systems primarily focus on users' interaction behaviors with videos, neglecting comment content and interaction in user preference modeling.
In this paper, we propose a novel recommendation approach called \textbf{LSVCR} that utilizes user interaction histories with both \emph{videos} and \emph{comments} to jointly perform personalized video and comment recommendation. 
Specifically, our approach comprises two key components: sequential recommendation~(SR) model and supplemental large language model~(LLM) recommender.
The SR model functions as the primary recommendation backbone (\emph{retained in deployment}) of our method for efficient user preference modeling.
Concurrently, we employ a LLM as the supplemental recommender (\emph{discarded in deployment}) to better capture underlying user preferences derived from heterogeneous interaction behaviors.
In order to integrate the strengths of the SR model and the supplemental LLM recommender, we introduce a two-stage training paradigm.
The first stage, personalized preference alignment, aims to align the preference representations from both components, thereby enhancing the semantics of the SR model.
The second stage, recommendation-oriented fine-tuning, involves fine-tuning the alignment-enhanced SR model according to specific objectives.
Extensive experiments in both video and comment recommendation tasks demonstrate the effectiveness of LSVCR. 
Moreover, online A/B testing on \emph{KuaiShou} platform verifies the practical benefits of our approach. 
In particular, we attain a cumulative gain of 4.13\% in comment watch time.

\end{abstract}



\sloppy
\maketitle

\section{Introduction}
\label{sec:intro}
Nowadays, recommender systems play an essential role in alleviating information overload by providing personalized recommendation services with high-quality content resources.  
As a typical application scenario,  online video platforms (\eg \emph{YouTube}, \emph{TikTok}, \emph{KuaiShou}) employ recommender systems to deliver customized videos to users, in which they primarily concentrate on capturing users' video watching interests based on user-video interaction logs~\cite{DBLP:conf/uai/RendleFGS09,gru4rec,sasrec,DBLP:conf/kdd/LinWMZWJW22}.
With the growth of online video communities, the textual comments posted to the videos become increasingly crucial for enhancing the overall watching experience of users, providing complement or divergence information compared to the original video content.
Our statistics show that more than 60\% of users on our video platform regularly view corresponding comments when watching the videos and express their interests through comments.
In light of this, this paper aims to leverage both video and comment data to improve the recommendation quality and user engagement. 

\begin{figure}[]
\centering
\includegraphics[width=0.98\linewidth]{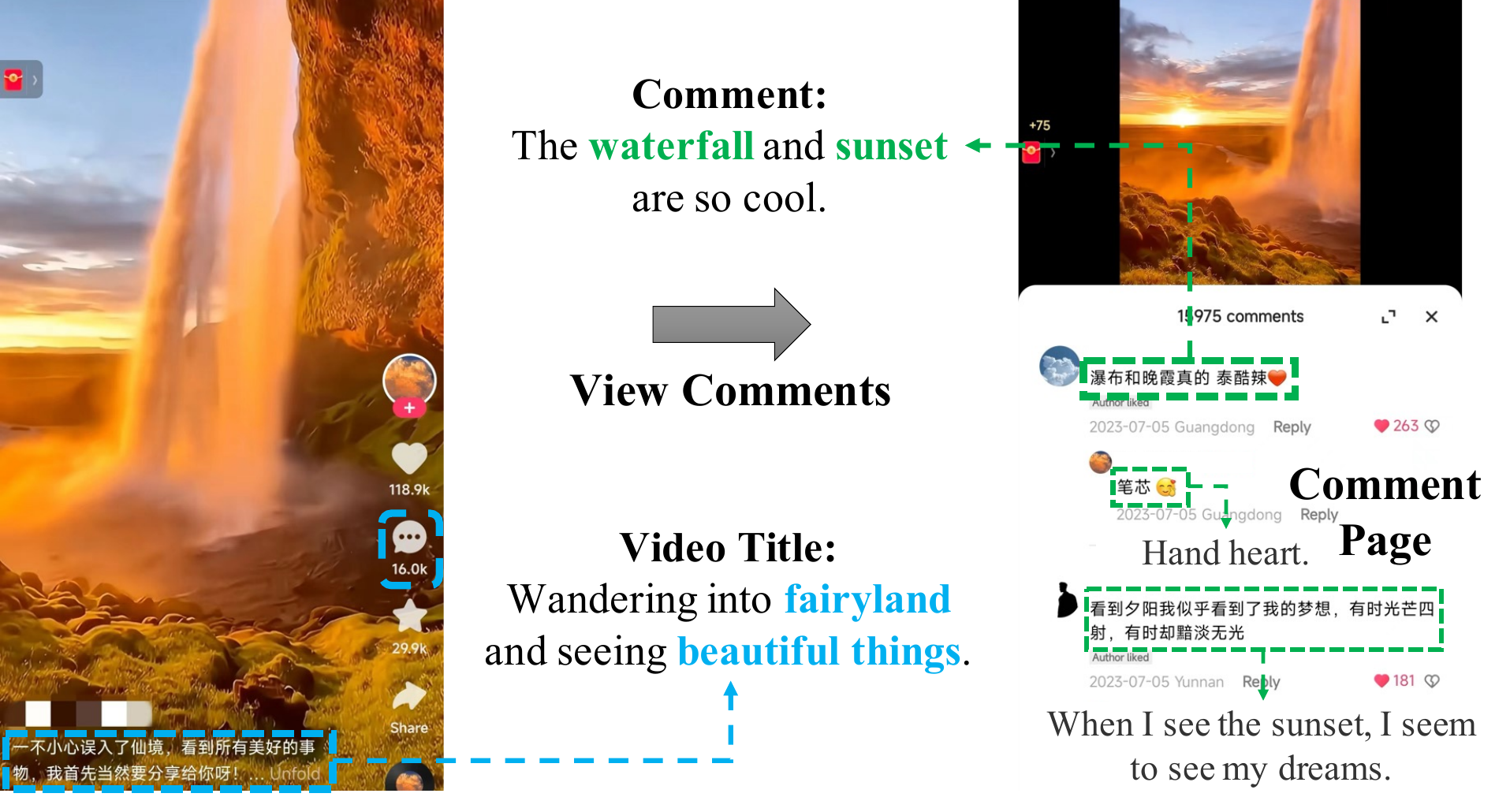}
\captionsetup{skip=3pt}
\caption{A snapshot of one micro-video and its corresponding comment area on \emph{KuaiShou} platform.}
\vspace{-10pt}
\label{fig:intro}
\end{figure}

To be more precise, we propose to study the joint tasks of video and comment recommendation in the specific scenario of a micro-video platform (\eg \emph{KuaiShou}, \emph{TikTok} ).
Figure~\ref{fig:intro} presents a snapshot of a micro-video and its associated comments on our platform. 
When a user is watching a video and clicks the ``\emph{comment}'' button, the comment page below the video will unfold, displaying a list of comments to the user. 
To develop a comprehensive solution, we integrate the modeling of videos and comments into a unified framework, considering the following three aspects:
(1) While comments provide supplemental textual information for a video, their semantic modeling should be grounded in the video content. For example, in Figure~\ref{fig:intro}, the keywords ``\emph{sunset}'' and ``\emph{waterfall}'' in the comment complement the video features and should be understood within the context of ``\emph{fairyland}'' and  ``\emph{beautiful things}''. 
(2) User's interactions and feedback on comments indicate potential interest in watching the corresponding video. When a user actively interacts with the comments of a particular video, he/she is more likely to enjoy videos with similar content or comments.
(3) In our video platform, a user could make multiple interactions with videos (\eg  \emph{like},\emph{collect}, \emph{share}) and corresponding comments (\emph{like}, \emph{reply}). Both interaction histories can capture users' personalized video and comment preferences, which might suggest delivering high-quality video and customized comments instead of monotonous hot comments.

Recently, large language models~(LLMs) have demonstrated great potential in recommender systems~\cite{DBLP:journals/corr/abs-2205-08084,DBLP:conf/emnlp/WangTZWW23,DBLP:journals/corr/abs-2311-09049,DBLP:conf/recsys/BaoZZWF023},  due to their excellent semantic understanding and knowledge reasoning capabilities. 
In our task setting, we have abundant textual input such as \emph{video title} and \emph{comment content}, which naturally motivates us to leverage the impressive semantic modeling capability of LLMs for developing our approach.  
Another potential advantage of using LLMs in our setting is that the interaction histories of videos and comments essentially provide heterogeneous expressions of user behaviors. 
Thus, we can leverage LLMs to reason about the underlying user preferences from the two kinds of interactions.  
However, the unaffordable computation costs of LLMs restrict their deployment as the online recommender in large-scale industrial systems~\cite{DBLP:journals/corr/abs-2305-07001,DBLP:journals/corr/abs-2305-08845,xu2024prompting}.
Considering this issue, we only adopt a LLM in training as a supplement to enhance the semantics of our recommendation backbone.

To this end, in this paper, we propose a novel framework that consolidates the benefits of \textbf{L}LM with the \textbf{S}equential recommendation model for joint  \textbf{V}ideo and \textbf{C}omment \textbf{R}ecommendation, namely \textbf{LSVCR}.
Our framework consists of two key components: conventional sequential recommendation~(SR) model and supplemental LLM recommender. 
To begin with, we develop our recommendation backbone (\emph{retained in deployment}) based on the SR model to achieve efficient user preference modeling.
Furthermore, we introduce a supplemental LLM recommender (\emph{discarded in deployment}) to enhance the preference semantics of the SR model.
By verbalizing both types of user interaction histories in the format of textual instructions, we aim to leverage the LLM's semantic understanding and knowledge reasoning capabilities for enhanced preference modeling.
In order to integrate the merits of the above two components, we design a training paradigm consisting of two stages: personalized preference alignment and recommendation-oriented fine-tuning.
For the alignment stage, we jointly learn the video and comment recommendation tasks while aligning the preference representation of the SR model with that of the supplemental LLM recommender.
For the fine-tuning stage, we discard the supplemental LLM recommender and further adapt the alignment-enhanced SR model to specific tasks for better recommendation performance.

To evaluate our approach, we conduct extensive experiments on a large real-world industrial dataset and a public Amazon benchmark. 
The experimental results demonstrate the significant effectiveness of LSVCR compared to competitive baselines for both video and comment recommendation tasks.
Additionally, online A/B testing verifies the benefits of LSVCR in the industrial recommender system. 
Especially in the comment end, we achieve a 4.13\% increase in watch time and a 1.36\% gain in interaction number.

\section{Related Work}

\paratitle{Sequential Recommendation.}
Sequential recommendation aims to mine users' personalized preferences and recommend the next item for each user. This field has attracted a lot of studies due to its advantages in capturing the dynamic characteristics of user behaviors~\cite{gru4rec,sasrec,bert4rec}.
Early methods follow the Markov Chain assumption and apply matrix factorization to model item-item transfer relationships~\cite{DBLP:conf/www/RendleFS10,DBLP:conf/icdm/HeM16}.
Recently, typical works are mostly based on various deep neural networks to better user behavior modeling, including RNN~\cite{gru4rec,DBLP:conf/cikm/LiRCRLM17}, CNN~\cite{DBLP:conf/wsdm/TangW18,DBLP:conf/wsdm/YuanKAJ019}, GNN~\cite{DBLP:conf/aaai/WuT0WXT19,DBLP:conf/ijcai/XuZLSXZFZ19}, and Transformer~\cite{sasrec,bert4rec}.
Furthermore, several studies focus on utilizing additional contextual information (\eg title, description, category) to enhance item sequence modeling~\cite{DBLP:conf/ijcai/ZhangZLSXWLZ19,DBLP:conf/cikm/ZhouWZZWZWW20,DBLP:conf/sigir/XieZK22}.
Unlike these works, this paper specifically concentrates on video comments, incorporating abundant textual information and user feedback from comment pages into sequential recommendation.

\paratitle{Text-Enhanced Recommendation.}
On various application platforms, items are associated with abundant textual features, which motivates researchers to engage in text-enhanced recommendation models. 
Recently, large language models~(LLMs) have gained significant popularity and shown great potential in wide-ranging fields~\cite{DBLP:journals/corr/abs-2303-18223, DBLP:conf/emnlp/JiangZDYZW23,DBLP:journals/corr/abs-2305-19860}. 
In the realm of text-enhanced recommendation, LLM generally plays distinct roles:
(1) LLM serves as a text feature encoder, allowing the text features to be encoded asynchronously. Subsequently, a conventional recommendation model utilizes knowledge-aware text embeddings for recommendation~\cite{DBLP:conf/sigir/WuWQ021,DBLP:conf/kdd/HouMZLDW22,DBLP:conf/sigir/YuanYSLFYPN23,DBLP:conf/recsys/HarteZLKJF23,DBLP:journals/corr/abs-2310-15950,vqrec}. 
Although effective, this approach often requires special mechanisms to enhance text feature modeling for better performance, such as contrastive learning~\cite{DBLP:conf/kdd/HouMZLDW22,DBLP:journals/corr/abs-2310-15950}.
(2) LLM serves as a recommender. This approach typically constructs the user's interaction history and other contexts into textual instructions, and then LLM is instructed to directly generate recommendation targets~\cite{DBLP:journals/corr/abs-2305-07001,DBLP:journals/corr/abs-2305-08845,xu2024prompting,DBLP:journals/corr/abs-2311-09049,DBLP:conf/recsys/BaoZZWF023}.
More recently, several works leverage various methods to improve the adaptation of LLM for recommendation scenarios, including incorporating collaborative embedding~\cite{DBLP:journals/corr/abs-2310-19488,DBLP:journals/corr/abs-2312-02443,DBLP:journals/corr/abs-2312-02445}, and semantic alignment~\cite{DBLP:journals/corr/abs-2311-01343,DBLP:journals/corr/abs-2311-09049}. 
These methods fully utilize the excellent capabilities of LLM and achieve astonishing results, but they still cannot avoid the high computational cost and slow inference time caused by LLM.
As a comparison, we aim to consolidate the merits of these two approaches through personalized preference alignment in order to achieve effective textual semantics understanding and heterogeneous interaction behavior modeling.

\begin{figure*}[]
\centering
\includegraphics[width=0.99\linewidth]{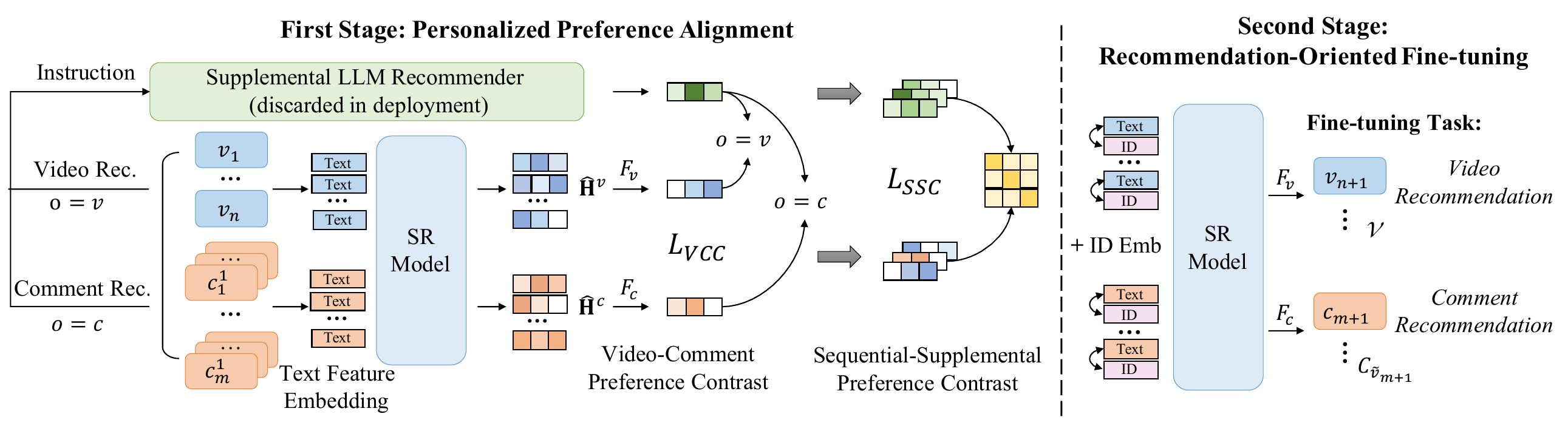}
\caption{The overall framework of our LSVCR. $v_i$ and $c_j^k$ denote the video and comment.
$o \in \{v, c\}$ represents the instruction/task objective (video or comment recommendation) of each data instance.
``SR model'' is the short of \underline{S}equential \underline{R}ecommendation Model.
$F_v$ and $F_c$ denote the preference extraction function corresponding to each task in Eq.~\eqref{eq:sv} and Eq.~\eqref{eq:sc}.}
\label{fig:model}
\end{figure*}

\section{Problem Formulation}
\label{sec:prob}
In our task, we consider the online micro-video scenario, where users can freely watch interesting videos and further view comments below specific videos. 
Different from previous video recommendation studies~\cite{DBLP:conf/uai/RendleFGS09,gru4rec,sasrec,DBLP:conf/kdd/LinWMZWJW22}, we consider both \emph{video recommendation} and \emph{comment recommendation}, in which comment recommendation is helpful to enhance the user engagement and improve the user experience. 
Formally, we denote the sets of users and videos as $\mathcal{U}$ and $\mathcal{V}$ respectively. 
Each video $v \in \mathcal{V}$ is associated with a textual title and a list of corresponding comments, denoted as $\mathcal{C}_{v}$. 
For a user $u \in \mathcal{U}$, there exist two chronological interaction sequences: (1) \emph{Video interactions}  $\mathcal{S}_v = [v_1, v_2,\dots, v_n]$, where $v_i$ is the $i$-th watched video by the user.
(2) \emph{Comment interactions}  $\mathcal{S}_c = [r_1,r_2, \dots, r_m]$,
where $r_j$ is the $j$-th comment interaction record of the user, and each record contains a set of comments that the user has interacted with below video $\tilde{v}_j$, denoted as $(c_j^1,\dots,c_j^k)$
In addition, we denote the original videos of these comments as $\mathcal{S}_{\tilde{v}} = [\tilde{v}_1, \tilde{v}_2,\dots, \tilde{v}_m]$ for further elaboration. 
In our platform, a user can only engage in the comment thread without interacting with the corresponding video.
Thus, the videos in $\mathcal{S}_v$ and $\mathcal{S}_{\tilde{v}}$ may not be the same, but likely share overlapped ones.  

Based on the above notations, we focus on the tasks of sequential video and comment recommendation. 
Regarding video recommendation, given the video interaction history $\mathcal{S}_v$ and the auxiliary comment interaction history $\mathcal{S}_c$, the task aims to predict the next video that the user may be interested in. 
We attempt to enhance user preference modeling with feedback on video comments.
For comment recommendation, the primary focus is on $\mathcal{S}_c$, with $\mathcal{S}_v$ serving as a supplement. 
When the user is watching video $\tilde{v}_{m+1}$ and clicks the ``comment''` button, the task objective is to predict the comments that the user may interact with for the current video and display personalized comment content to the user.

\section{Methodology}
In this section, we present our proposed \textbf{LSVCR} for joint video and comment recommendation. Its overall framework is depicted in Figure~\ref{fig:model}. 
We first introduce the SR model for sequential user preference modeling in Section~\ref{sec:srmodel}.
Then we incorporate the supplemental LLM recommender in Section~\ref{sec:llmmmodel} to enhance preference modeling.
After that, a personalized preference alignment approach is proposed in Section~\ref{sec:pretrain} to align the SR model with the supplemental LLM recommender.
Finally, we fine-tune the alignment-enhanced SR model for video or comment recommendation tasks in Section~\ref{sec:finetune}.


\subsection{Sequential User Preference Modeling}\label{sec:srmodel}
As we mentioned in Section~\ref{sec:intro}, it is resource-intensive to directly employ LLM as a recommender for online service~\cite{xu2024prompting}.
Thus, we consider a more economical way to utilize LLM as an offline text feature encoder and utilize a sequential recommendation model~(called SR model) as the backbone for user behavior modeling~\cite{DBLP:conf/sigir/YuanYSLFYPN23,DBLP:conf/kdd/HouMZLDW22}.

\subsubsection{Text Feature Embedding with LLM}
Following our problem formulation,  we use the user's interaction sequences, \ie $\mathcal{S}_v$ and $\mathcal{S}_c$, to construct inputs.
First, we employ LLM to encode the titles and comments involved in the inputs. 
Then, we combine the outputs as the representation of each interaction, which can be written as:
\begin{align}
\mathbf{z}^{v}_i & = [\operatorname{LLM}(t_i)||\operatorname{LLM}(c_i)]\mathbf{W}_1, \label{eq:zv}\\
\mathbf{z}^{c}_j & = [\operatorname{LLM}(t_{j})||\operatorname{MEAN}\big(\operatorname{LLM}(c_{j}^1),\dots,\operatorname{LLM}(c_{j}^k)\big)]\mathbf{W}_1,
 \label{eq:zc}
\end{align}
where $t_i$ and $c_i$ denote the title and a hot comment of video $v_i$ in $\mathcal{S}_v$.
$t_j$ denotes the title of video $\tilde{v}_j$ corresponding to the interaction record $r_j$ in $\mathcal{S}_c$.
``||'' denotes concatenation operation,  $\mathbf{W}_1 \in \mathbb{R}^{2d \times d}$ is a trainable parameter for linear projection, and 
$\operatorname{MEAN}(\cdot)$ denotes mean pooling. 
$\mathbf{z}^{v}_i$ and $\mathbf{z}^{c}_j$ represent the text feature embeddings for one video/comment interaction, and the text feature embeddings of interaction sequences are denoted as $\mathbf{Z}^{v} \in \mathbb{R}^{n \times d}$ and $\mathbf{Z}^{c} \in \mathbb{R}^{m \times d}$.

\subsubsection{Randomized Positional Encoding for Length Extension}
\label{sec:randposi}
To model the sequential pattern of user histories, we maintain two learnable positional encoding matrices, denoted as $\mathbf{P}^v \in \mathbb{R}^{L_v \times d}$ and $\mathbf{P}^c \in \mathbb{R}^{L_c \times d}$, respectively. 
$L_v$ and $L_c$ are the maximum lengths of the interaction sequences.
In conventional sequential recommendation, the length can generally be set longer~\cite{gru4rec,sasrec}.
However, in the personalized preference alignment stage, it is set shorter, limited by LLM input length and computational consumption.
Inspired by the length extension technique in the NLP~\cite{DBLP:conf/acl/RuossDGGCBLV23}, we randomly select an ordered subset of positional encodings during the alignment stage, while applying the entire set in fine-tuning and inference processes. 
Formally, the positional encodings of $\mathcal{S}_v^u$ are sampled as follows:
\begin{align}
\widetilde{\mathbf{P}}^v &= [\mathbf{p}_{i_1}, \mathbf{p}_{i_2}, \dots, \mathbf{p}_{i_{n}}],
\end{align}
where $\mathbf{p}_{i_k}$ denotes the $i_k$-th positional encoding in $\mathbf{P}^v$. Each index $i_k$ is sampled from $\{1, \dots, L_v\}$ without replacement, and all sampled indices are arranged in ascending order. 
Additionally, the positional encodings $\widetilde{\mathbf{P}}^c$ for $\mathcal{S}_c^u$ are sampled using a similar process. 

\subsubsection{Sequence Representation Learning}
\label{sec:seq_encode}
Given the user interaction histories and sampled position encodings, we use the Transformer backbone~\cite{DBLP:conf/nips/VaswaniSPUJGKP17} to model video and comment interaction sequences. 
The above sequence embeddings (\ie $\mathbf{Z}^{v}$ and $\mathbf{Z}^{c}$) and positional encodings (\ie $\widetilde{\mathbf{P}}^v$ and $\widetilde{\mathbf{P}}^c$) are summed up as inputs of Transformer layers. 
The calculations can be formally written as:
\begin{align}
\mathbf{H}^v =\operatorname{TRM}^v(\mathbf{Z}^{v} + \widetilde{\mathbf{P}}^v), \ \ \ \ 
\mathbf{H}^c =\operatorname{TRM}^c(\mathbf{Z}^{c} + \widetilde{\mathbf{P}}^c),
\label{eq:enc}
\end{align}
where $\mathbf{H}^v \in \mathbb{R}^{n \times d}$ and $\mathbf{H}^c \in \mathbb{R}^{m \times d}$ are sequence representations of user video and comment interaction histories.

Next, a pair of fusion encoders based on multi-head attention (denoted by $\operatorname{MHA}(\mathbf{Q}, \mathbf{K}, \mathbf{V})$) are used to achieve the cross-fusion between two user historical sequences:
\begin{align}
\widetilde{\mathbf{H}}^v = \operatorname{MHA}^v(\mathbf{\mathbf{H}^v}, \mathbf{\mathbf{H}^c}, \mathbf{\mathbf{H}^c}), &\quad 
\widehat{\mathbf{H}}^v = [\mathbf{H}^v||\widetilde{\mathbf{H}}^v] \mathbf{W}_2, 
\label{eq:vfus}\\
\widetilde{\mathbf{H}}^c = \operatorname{MHA}^c(\mathbf{\mathbf{H}^c}, \mathbf{\mathbf{H}^v}, \mathbf{\mathbf{H}^v}), &\quad 
\widehat{\mathbf{H}}^c = [\mathbf{H}^c||\widetilde{\mathbf{H}}^c] \mathbf{W}_3, 
\label{eq:cfus}\
\end{align}
where $\mathbf{W}_2 \in \mathbb{R}^{2d \times d}$ and $\mathbf{W}_3 \in \mathbb{R}^{2d \times d}$ are learnable parameters for representation fusion, and 
$\widehat{\mathbf{H}}^v \in \mathbb{R}^{n \times d}$ and $\widehat{\mathbf{H}}^c \in \mathbb{R}^{m \times d}$ represent the fused video and comment sequence representations.

\subsubsection{Preference Extraction from Sequence Representations}
\label{sec:extra}
For the tasks of video and comment recommendation, we employ different methods to extract user preferences from sequence representations. 
Specifically, we exploit additive attention~\cite{DBLP:journals/corr/BahdanauCB14} in video recommendation to manage the importance of different representations:
\begin{align}
\mathbf{s}^v &= F_v(\widehat{\mathbf{H}}^v) = \sum_{i=1}^{n} \alpha_i \mathbf{h}_i^v, \ \ \ \  
\alpha_i &= \frac{\text{exp}(f(\mathbf{h}_i^v))}{\sum_{k=1}^{n} \text{exp}(f(\mathbf{h}_k^v))},
\label{eq:sv}
\end{align}
where $\mathbf{h}_i^v$ is the $i$-th vector of the matrix $\widehat{\mathbf{H}}^v$, $f$ is an attention weight function implemented as a multi-layer perceptron (MLP), and $\mathbf{s}^v$ denotes the representation of user video preference.

As for the comment recommendation task, which involves the video that the user is watching, we regard the title embedding (denoted as $\mathbf{e}^t_{m+1} = \operatorname{LLM}(t_{m+1})$) of the current video as a query for representation aggregation:
\begin{align}
\mathbf{s}^c &= F_c(\widehat{\mathbf{H}}^c) = \sum_{j=1}^{m} \beta_j \mathbf{h}_j^c, \ \ \ \ 
\beta_j &= \frac{\text{exp}(g(\mathbf{e}^t_{m+1},\mathbf{h}_j^c))}{\sum_{k=1}^{m} \text{exp}(g(\mathbf{e}^t_{m+1},\mathbf{h}_k^c))}.
\label{eq:sc}
\end{align}
where $g(\cdot, \cdot)$ is an attention weight function (similar to the attention function in Transformer) with $\mathbf{e}^t_{m+1}$ as the query, $\mathbf{s}^c$ denotes the representation of user comment preference on video $\tilde{v}_{m+1}$.

\subsection{Enhanced Preference Modeling via LLM}
\label{sec:llmmmodel}
In recommendation scenarios, LLM generally plays two roles: as a text feature encoder (Section~\ref{sec:srmodel}) or a recommender~\cite{DBLP:journals/corr/abs-2305-07001,DBLP:conf/recsys/BaoZZWF023}. 
The former offers greater flexibility and efficiency, while the latter is beneficial for maximizing LLM’s excellent semantic understanding and knowledge reasoning capabilities. 
Therefore, we further introduce enhanced user preference modeling through a LLM recommender to improve the semantics of the SR model.
Notably, this component is only included \emph{in the alignment training stage} and does not exist in the subsequent fine-tuning and practical deployment.  

Following previous works~\cite{DBLP:journals/corr/abs-2305-08845,DBLP:conf/recsys/DaiSZYSXS0X23,DBLP:journals/corr/abs-2305-07001}, we format the user interaction behaviors in the form of textual instructions and utilize LLM for generative recommendation.
Specifically, the instruction templates for two tasks are as follows:
\aptLtoX{\begin{quote}

\textbf{Video Recommendation:} \emph{A user has interacted with the following videos:} [\emph{VideoInterList}]\emph{, and his/her comment interaction history is as follows:} [\emph{CommentInterList}]\emph{. Please recommend the next video for him/her based on these interaction histories:} 

\textbf{Comment Recommendation:} \emph{A user's comment interaction history is as follows:} [\emph{CommentInterList}]\emph{, and he/she has interacted with the following videos:} [\emph{VideoInterList}]\emph{. Now he/she is watching a video:} [\emph{VideoTitle}]\emph{. Please generate a potentially interactive comment based on the above historical interactions and the current video content:} 

\end{quote}}{\begin{quote}
\begin{adjustwidth}{-7mm}{-7mm}

\textbf{Video Recommendation:} \emph{A user has interacted with the following videos:} [\emph{VideoInterList}]\emph{, and his/her comment interaction history is as follows:} [\emph{CommentInterList}]\emph{. Please recommend the next video for him/her based on these interaction histories:} 

\textbf{Comment Recommendation:} \emph{A user's comment interaction history is as follows:} [\emph{CommentInterList}]\emph{, and he/she has interacted with the following videos:} [\emph{VideoInterList}]\emph{. Now he/she is watching a video:} [\emph{VideoTitle}]\emph{. Please generate a potentially interactive comment based on the above historical interactions and the current video content:} 

\end{adjustwidth}
\end{quote}}

In the instruction templates, 
``[\emph{VideoInterList}]'' represents a list of video interactions, where each record contains a video title and a popular comment, \eg  ``Title: \emph{Review of the 2022 Beijing Winter Olympics}; Popular comment: \emph{Figure skating is so beautiful!}''. As can be seen from this example, the associated comment provides additional illustration for this video, with a special focus on figure skating. 
``[\emph{CommentInterList}]'' represents a list of comment interactions, where each record contains the comments that the user has interacted with and the corresponding video title.
For the video recommendation task, LLM is instructed to generate the title of the next video. 
As for the comment recommendation task, it is based on the video that the user is watching (\ie ``[\emph{VideoTitle}]''). LLM is instructed to generate the comment that the user may interact with.
Finally, we denote the instruction as $I(\mathcal{S}_v, \mathcal{S}_c)$ and feed it into LLM: 
\begin{align}
\tilde{\mathbf{s}} = \text{LLM}(I(\mathcal{S}_v, \mathcal{S}_c))[-1],
\label{eq:llms}
\end{align}
where $[-1]$ means taking the hidden state of the final token, $\tilde{\mathbf{s}}$ is the enhanced preference representation from the LLM recommender.

\subsection{Personalized Preference Alignment}
\label{sec:pretrain}
In our approach, the SR model and the supplemental LLM recommender learn user preference representations from different perceptions.
To enhance the SR model by incorporating more informative preference knowledge from the LLM recommender, we introduce personalized preference alignment as the first stage of training.
Specifically, we propose two alignment perspectives: sequential-supplemental preference contrast and video-comment preference contrast, to align preference semantics from both components.

\subsubsection{Sequential-Supplemental Preference Contrast} 
Contrastive learning is typically used to align instance representations across different latent spaces~\cite{DBLP:conf/icml/ChenK0H20,DBLP:conf/emnlp/GaoYC21}. 
In our LSVCR, the primary objective is to achieve alignment of personalized preference representations between the SR model and the supplemental LLM recommender. 
Specifically, we first perform data augmentation via comment diversification on inputs and then employ InfoNCE~\cite{DBLP:journals/jmlr/GutmannH10} with in-batch negatives to align the output representations of two models.

\paratitle{Input augmentation via comment diversification.}
\label{sec:comdiv}
In our context, apart from the comments that users have interacted with, there is also a vast amount of comment data. To improve data utilization, we introduce comment diversification for data augmentation.
Specifically, we apply hot comments to augment user video and comment interaction histories with a certain probability:
(1) For each video interaction record, we select two distinct popular comments to pair with the same video as inputs for the SR model and the LLM recommender. 
(2) For a given comment interaction record, assume that it contains k comments that the user has interacted with. We first sample k popular comments under the video as virtual interaction comments and mix them with the actual interaction comments.
After that, we randomly allocate mixed interaction comments to two groups, each containing k comments. These two groups will be used to construct inputs for the SR model and the LLM recommender respectively.

\paratitle{Preference representation contrastive learning.}
In the personalized preference alignment stage, we jointly learn video and comment recommendation tasks. 
For each data instance, the sequential preference representation of the SR model is calculated by Eq.~\eqref{eq:sv} or Eq.~\eqref{eq:sc} according to the task objective.
To simplify, here we no longer distinguish between the two tasks, and the representations are uniformly denoted as $\mathbf{s}$.
Additionally, by the approach mentioned in Section~\ref{sec:llmmmodel}, we obtain the enhanced preference representation of the LLM recommender~(different task instances use different instruction templates), denoted as $\tilde{\mathbf{s}}$.
The sequential-supplemental preference contrastive loss can be formulated as:
\begin{align}
\mathcal{L}_{SSC} & = \frac{1}{2} \left( \operatorname{InfoNCE}(\mathbf{s},\tilde{\mathbf{s}},\mathcal{R}_{\tilde{\mathbf{s}}}) + \operatorname{InfoNCE}(\tilde{\mathbf{s}},\mathbf{s}, \mathcal{R}_{\mathbf{s}}) \right),
\label{eq:sscloss}
\end{align}
where $\mathcal{R}_{\mathbf{s}}$ and $\mathcal{R}_{\tilde{\mathbf{s}}}$ denote batch preference representations that are generated by the SR model and the LLM recommender, respectively. $\operatorname{InfoNCE}(\cdot,\cdot,\cdot)$ represents InfoNCE loss, which can be written as:
\begin{align}
\operatorname{InfoNCE}(\mathbf{x},\mathbf{y}^{+}, \mathcal{O}_{\mathbf{y}}) = - \log \frac{\text{exp}(\text{cos}(\mathbf{x},\mathbf{y}^{+}) / \tau)}{\sum_{\mathbf{y} \in \mathcal{O}_{\mathbf{y}} } \text{exp}(\text{cos}(\mathbf{x},\mathbf{y}) / \tau)},
\end{align}
where $\mathbf{x}$ and $\mathbf{y}^{+}$ denote a pair of positive instances for contrastive learning.  $\mathcal{O}_{\mathbf{y}}$ denotes contrastive instances consisting of both positive and negative samples. 
$\tau$ is a temperature coefficient.

\subsubsection{Video-Comment Preference Contrast}
\label{sec:vcc}
Furthermore, we disentangle user preferences derived from video and comment interaction sequences, based on the enhanced preference representation provided by the LLM recommender.
The core idea is that when LLM is instructed to make video recommendations, the preference representation (\ie $\tilde{\mathbf{s}}$) should be aligned with that extracted from the video sequence (\ie $\widehat{\mathbf{H}}^v$), rather than from the comment sequence (\ie $\widehat{\mathbf{H}}^c$), and vice versa.
Formally, we denote the task objective (video or comment recommendation) of an instance as $o \in \{v, c\}$.
The video-comment preference contrastive loss is as follows:
\begin{align}
\mathcal{L}_{VCC}
& = - \log \frac{\text{exp}(\text{cos}(\tilde{\mathbf{s}}, F_{o} (\widehat{\mathbf{H}}^{o})) / \tau)}{\sum_{\tilde{o} \in \{v, c\}} \text{exp}(\text{cos}(\tilde{\mathbf{s}}, F_{o}(\widehat{\mathbf{H}}^{\tilde{o}}) ) /\tau)},
\label{eq:vccloss}
\end{align}
where $F_{o}$ denotes the preference extraction function corresponding to each instance in Eq.~\eqref{eq:sv} (\ie $F_v$) or Eq.~\eqref{eq:sc} (\ie $F_c$).

\subsubsection{Alignment Training Objective}
In our approach, we consider the next video title as the target text for video recommendation and the comment that the user may interact with as the target text for comment recommendation.
LLM is trained by following a typical procedure of instruction tuning~\cite{DBLP:conf/iclr/WeiBZGYLDDL22}:
\begin{align}
\mathcal{L}_{LM} = - \sum_{i=1}^{|T|}\operatorname{log}P(T_{i}|I,T_{<i}),
\label{eq:llmloss}
\end{align}
where $I$ and $T$ represent the instruction and target text, $T_{i}$ is the $i$-th token of $T$, and $T_{<i}$ denotes the tokens before $T_{i}$.
In terms of the SR model, we harness the following loss for optimization:
\begin{align}
\mathcal{L}_{SR}
& = \operatorname{InfoNCE}(\mathbf{s},\mathbf{e}, \mathcal{T}_{\mathbf{e}})
\end{align}
where $e = \operatorname{LLM}(T)$, and $\mathcal{T}_{\mathbf{e}}$ denotes a batch of target embeddings.

In the end, the overall loss of personalized preference alignment, which combines the above objectives, is as follows:
\begin{align}
    \mathcal{L}_{ALI} = \mathcal{L}_{LM} + \lambda \mathcal{L}_{SR} + \mu (\mathcal{L}_{SSC} + \mathcal{L}_{VCC}),
    \label{eq:all_loss}
\end{align}
where $\lambda$ and $\mu$ are hyperparameters that balance various objectives.

\subsection{Recommendation-Oriented Fine-tuning} 
\label{sec:finetune}
After the personalized preference alignment stage, we further fine-tune the alignment-enhanced SR model for video and comment recommendation tasks respectively.
The supplemental LLM recommender will be discarded during the fine-tuning and deployment.
\subsubsection{Fine-tuning}
Since the first stage only utilizes textual features, we incorporate video ID embedding into our framework during the fine-tuning stage to enhance the overall performance.
The particular method is to add the corresponding ID embedding to the inputs of the SR model in Eq.~\eqref{eq:zv} and Eq.~\eqref{eq:zc}.
Moreover, to ensure representation compatibility between ID and text embeddings, we introduce the following ID-text regularization loss:
\begin{align}
\mathcal{L}_{REG}
& = \operatorname{InfoNCE}(\mathbf{e}_i^{id},\mathbf{z}_i^{v},\mathcal{Z}_i^v )  +  \operatorname{InfoNCE}(\mathbf{e}_{j}^{id},\mathbf{z}_j^{c},\mathcal{Z}_j^c ),\label{eq:regloss}
\end{align}
where $\mathbf{e}_i^{id}$ and $\mathbf{e}_{j}^{id}$ denote the ID embeddings of video $v_i$ in $\mathcal{S}_v$ and $\tilde{v}_j$ in $\mathcal{S}_{\tilde{v}}$, respectively. 
The sets $\mathcal{Z}_i^v$ and $\mathcal{Z}_j^c$ consist of both the positive and negative text embeddings sampled from the batch data.

For the video recommendation task, our objective is similar to target text contrastive learning, but we use all videos as candidates instead of in-batch negatives. The loss can be written as:
\begin{align}
\mathcal{L}_{VR}
& = \operatorname{InfoNCE}(\mathbf{s}^{v},(\mathbf{e}^{id}_{n+1} + \mathbf{e}^t_{n+1}), \mathcal{V}),
\end{align}
where $\mathcal{V}$ is the entire set of candidate videos. $\mathbf{e}^t_{n+1} = \operatorname{LLM}(t_{n+1})$ represents the title embedding of the next video $v_{n+1}$. 
Finally, our overall optimization loss function for video recommendation in the fine-tuning stage can be denoted as follows:
\begin{align}
    \mathcal{L}_{VFT} = \mathcal{L}_{VR} + \eta \mathcal{L}_{REG},
\end{align}
where $\eta$ is the hyperparameter for the regularization loss.

Regarding the comment recommendation task, we employ $\mathbf{s}^c$~(Eq.~\eqref{eq:sc}) for prediction, and all candidates (\ie $\mathcal{C}_{\tilde{v}_{m+1}}$) are selected from comments of video $\tilde{v}_{m+1}$ that the user is watching. Consequently, the objective of comment recommendation is as follows: 
\begin{align}
\mathcal{L}_{CR}= \operatorname{InfoNCE}(\mathbf{s}^{c}, \mathbf{e}^{c}_{m+1}, \mathcal{C}_{\tilde{v}_{m+1}}), 
\end{align}
where $\mathbf{e}^{c}_{m+1} = \operatorname{LLM}(c_{m+1})$ denotes the positive comment embedding that encoded by LLM. After combining with the regularization loss, we can also obtain the loss for comment recommendation during fine-tuning, denoted as $\mathcal{L}_{VFT}$.

\subsubsection{Time Complexity Analysis}
In practical deployment, we retain the fine-tuned SR model for recommendation, and discard the supplemental LLM recommender.
Besides, the textual information associated with the videos and comments is encoded offline to be directly accessed during online inference.
Therefore, we just discuss the time complexity of the subsequent sequential user preference modeling process. 
In particular, the time consumption of the SR model primarily involves encoding the two interaction sequences of video and comments~(Eq.~\eqref{eq:enc}).
This results in a time consumption of $O(ln^2d+lm^2d)$, where $l$ denotes the number of Transformer layers, $d$ denotes the model dimension, $n$ and $m$ denote the lengths of user video and comment interaction histories, respectively.
In contrast, the time complexity of directly using LLM as a recommender is $O(LN^2D)$, where $L$ is the layer number, $H$ denotes the hidden state size, and $N$ is the length of the input text.
Since $L>>l$ and $H>>d$, this makes the deployment of the LLM recommender far more resource-intensive than our approach. 
Compared to the text-based method FSDA~\cite{DBLP:conf/ijcai/ZhangZLSXWLZ19} with a time complexity of $O(2ln^2d)$ and the multi-behavior method DMT~\cite{DBLP:conf/cikm/GuDWZLY20} with a time consumption of $O(ln^2d+lm^2d)$, our approach has a comparable time complexity.

\section{Experiment}
\subsection{Experiment Setup}

\subsubsection{Dataset}
We evaluate the proposed framework on the large-scale industrial dataset constructed from \emph{KuaiShou} platform for video and comment recommendation tasks\footnote{\url{https://github.com/lyingCS/KuaiComt.github.io}}~\cite{our_data}.  
To be specific, the dataset is built from the interaction logs of 19,691 users from October 24 to October 31, 2023. 
For data preprocessing, we split the user interactions into training, validation, and testing sets according to the timestamp. 
The data from the last two days is used for validation and testing respectively, while the remaining data is utilized to train our framework.
The detailed statistics of the dataset are summarized in Table \ref{tab:dataset_info}. 

\begin{table}[]
\centering
\caption{Statistics of the preprocessed dataset. ``V'' and ``C''  denote video and comment interactions, respectively.}
\label{tab:dataset_info}
\renewcommand\arraystretch{1.03}
\setlength{\tabcolsep}{0.8mm}{
\begin{tabular}{ccccc}
\hline
\#Users & \#Videos & \#Comments & \#Interactions-V & \#Interactions-C \\
\hline
19,691 & 100,772 & 14,163,262 & 789,986 & 100,593 \\
\hline
\end{tabular}}
\end{table}
\subsubsection{Baseline Models}
We compare LSVCR with the following various competitive baselines:

\noindent \textbf{Video recommendation.}
The baselines for video recommendation can be divided into three categories:
(1) Traditional sequential recommendation models:
\textbf{Caser}~\cite{DBLP:conf/wsdm/TangW18} employs CNN to model user preference through horizontal and vertical convolutional filters.
\textbf{GRU4Rec}~\cite{gru4rec} is a RNN-based model that utilizes GRU to encode historical behavior sequences. 
\textbf{SASRec}~\cite{sasrec} uses a unidirectional Transformer for next item prediction.
\textbf{BERT4Rec}~\cite{bert4rec} adopts a bidirectional Transformer to model the item sequence.
\textbf{NARM}~\cite{DBLP:conf/cikm/LiRCRLM17} combines the GRU with an attention mechanism to capture users' interests. 
(2) Multi-behavior models:
\textbf{DMT}~\cite{DBLP:conf/cikm/GuDWZLY20} exploits multiple Transformers with MoEs to model diverse user behavior sequences.
\textbf{MBHT}~\cite{DBLP:conf/kdd/YangHXLYL22} empowers the Transformer with a multi-behavior hypergraph to capture diverse multi-behavior dependencies.
(3) Text-enhanced models:
\textbf{FDSA}~\cite{DBLP:conf/ijcai/ZhangZLSXWLZ19} applies separate self-attention networks to model item-level and feature-level sequences.
\textbf{S$^3$-Rec}~\cite{DBLP:conf/cikm/ZhouWZZWZWW20} proposes pre-training with mutual information maximization to learn the correlation between items and attributes.
\textbf{UniSRec}~\cite{DBLP:conf/kdd/HouMZLDW22} incorporates textual item embeddings with a MoE adapter to learn universal representations. We use both item ID and text.
\textbf{VQ-Rec}~\cite{vqrec} learns vector-quantized representations for sequential recommendation.

\noindent \textbf{Comment recommendation.}
The baselines for comment recommendation include:
(1) Traditional recommendation model:
\textbf{DSSM}~\cite{DBLP:conf/cikm/HuangHGDAH13} uses dual encoders to separately encode user and candidate comments.
(2) Sequential modeling methods:
\textbf{GRU} intuitively utilizes GRU~\cite{DBLP:conf/emnlp/ChoMGBBSB14} to derive short- and long-term user representations~\cite{DBLP:conf/acl/AnWWZLX19, gru4rec, DBLP:conf/cikm/LiRCRLM17}.
\textbf{ATT} employs an additive attention~\cite{DBLP:conf/nips/VaswaniSPUJGKP17} mechanism to model the user comment interaction sequence~\cite{DBLP:conf/ijcai/WuWAHHX19}.
\textbf{MHA} adopts multi-head attention~\cite{DBLP:conf/nips/VaswaniSPUJGKP17} and additive attention to obtain the representation of the user comment interaction history~\cite{DBLP:conf/ijcai/WuWAHHX19, DBLP:conf/emnlp/WuWGQHX19}.
\textbf{UniSRec}$_C$~\cite{DBLP:conf/kdd/HouMZLDW22} mirrors the model structure of UniSRec for comment modeling
(3) Query-based models:
\textbf{ZAM}~\cite{DBLP:conf/cikm/AiHVC19} introduces a zero vector into the attention mechanism to dynamically control the influence of user interaction history on product search. 
\textbf{TEM}~\cite{DBLP:conf/sigir/BiAC20} uses Transformer to encode the query and user interaction sequence for more flexible personalization adaptation.
For all comment recommendation baselines, we adopt trainable BERT~\cite{DBLP:conf/naacl/DevlinCLT19} as their text encoder. Furthermore, we consider the title embedding of the video being watched by the user as the query for query-based models.

\begin{table}[]
\centering
\caption{Performance comparison of video recommendation. The best and second-best results are highlighted in bold and underlined font, respectively. ``*'' denotes that the improvements are significant at the level of 0.01 with paired $t$-test.}
\label{tab:res_rec}
\huge
\resizebox{1.0\linewidth}{!}{
\renewcommand\arraystretch{1.05}
\begin{tabular}{l|ccccc}
\hline
Methods & Recall@5 & Recall@10 & NDCG@5 & NDCG@10 & MRR \\
\hline
Caser    & 0.2230                       & 0.2886                        & 0.1601                     & 0.1813                      & 0.1481                  \\
GRU4Rec  & 0.2249                       & 0.2896                        & 0.1647                     & 0.1856                      & 0.1535                  \\
BERT4Rec & 0.2193                       & 0.2815                        & 0.1612                     & 0.1814                      & 0.1504                  \\
SASRec   & 0.2311                       & 0.2953                        & 0.1706                     & 0.1913                      & 0.1592                  \\
NARM     & 0.2372                       & 0.2989                        & 0.1742                     & 0.1942                      & 0.1617                  \\
\hline
DMT      & 0.2236                       & 0.2903                        & 0.1633                     & 0.1848                      & 0.1522                  \\
MBHT     & 0.2421                       & 0.3038                        & 0.1780                     & 0.1980                      & 0.1651                  \\
\hline
FDSA     & 0.2277                       & 0.2922                        & 0.1667                    & 0.1875                    & 0.1551                  \\
S$^3$-Rec    & 0.2226                      & 0.2899                        & 0.1619                     & 0.1837                      & 0.1509                  \\
UniSRec$_{t+ID}$  & \underline{0.2570}                       & 0.3159                        & \underline{0.1875}                     & 0.2135                      & 0.1737                  \\
VQ-Rec  & 0.2540                       & \underline{0.3182}                        & 0.1813                     & \underline{0.2150}                      & \underline{0.1759}                  \\
\hline
LSVCR    & \textbf{0.2719*}                       & \textbf{0.3322*}                        & \textbf{0.2037*}                     & \textbf{0.2233*}                      & \textbf{0.1893*}                 \\
\hline
\end{tabular}}
\end{table}

\begin{table}[]
\centering
\caption{Performance comparison on the industrial dataset for comment recommendation. ``*'' denotes that the improvements are significant at the level of 0.01 with paired $t$-test.}
\label{tab:res_com}
\huge
\resizebox{1.0\linewidth}{!}{
\renewcommand\arraystretch{1.05}
\begin{tabular}{l|ccccc}
\hline
Methods   & Recall@5 & Recall@10 & NDCG@5 & NDCG@10 & MRR \\
\hline
DSSM    & 0.2092                       & 0.3386                        & 0.1447                     & 0.1884                      & 0.1573                  \\
GRU   & 0.2386                       & 0.3708                        & 0.1644                     & 0.2094                      & 0.1750               \\
ATT    & 0.2284                       & 0.3653                        & 0.1537                     & 0.2004                      & 0.1634                  \\
MHA    & 0.2339                       & \underline{0.3768}                        & 0.1646                     & 0.2131                      & 0.1775                  \\
UniSRec$_{C}$  & 0.2488                       & 0.3715                        & 0.1811                     & 0.2229                      & 0.1946                  \\
\hline
ZAM     & 0.2516         & 0.3699          & 0.1879       & 0.2283        & 0.2031    \\
TEM     & \underline{0.2533}          & 0.3685              & \underline{0.1898}                     & \underline{0.2302}                     & \underline{0.2057}                 \\
\hline
LSVCR   & \textbf{0.2821*}                       & \textbf{0.3901*}                        & \textbf{0.2175*}                     & \textbf{0.2541*}                      & \textbf{0.2303*}            \\
\hline
\end{tabular}}
\end{table}

\subsubsection{Evaluation Settings}
To evaluate the performance of video and comment recommendation tasks, we adopt three widely used metrics Recall@$K$, NDCG@$K$, and MRR, where $K$ is set to 5 and 10. 
For a fair comparison, the maximum historical sequence length is uniformly set to 50 for all comparison methods. 
For video recommendation, we pair the ground-truth item with 99 randomly sampled items. 
Regarding comment recommendation, the candidate number is limited to 100.
If the number of comments below the current video is less than 100, all comments will be used.
Conversely, negative sampling will be conducted until there are 100 candidates.

\subsubsection{Implementation Details}
For the personalized preference alignment stage, we utilize ChatGLM3~\cite{DBLP:conf/acl/DuQLDQY022,DBLP:conf/iclr/ZengLDWL0YXZXTM23} as our LLM backbone and perform low-rank adaptation based on LoRA~\cite{DBLP:conf/iclr/HuSWALWWC22}. 
We employ the AdamW optimizer with a learning rate of 3e-4 for optimization.
With the application of data parallelism and gradient accumulation, the total batch size reaches 128.
The loss coefficients, $\lambda$ and $\mu$, are set to 1 and 0.5 respectively. 
The temperature coefficient $\tau$ is assigned a value of 0.07.
The maximum lengths of user video and comment interaction histories are set to 20 for the alignment stage and 50 for the fine-tuning stage.
Finally, we conduct a total of 5000 alignment steps.
For the recommendation-oriented fine-tuning stage, the embedding dimension is set to 64, and the learning rate is tuned in \{0.005, 0.003, 0.001, 0.0005, 0.0001\}, which are consistent with all baseline models.
The hyperparameter $\eta$ is tuned in \{0.1, 0.5 \}. 
More details can be found in our implementation code \url{https://github.com/RUCAIBox/LSVCR}.

\subsection{Overall Results}

\textbf{Video Recommendation.}
The overall results of the video recommendation task are shown in Table~\ref{tab:res_rec}. From the results, we have the following observations:
Intuitively incorporating heterogeneous comment interaction behaviors (\ie DMT) is not better than traditional sequential recommendation models. 
In contrast, the hypergraph-enhanced Transformer (\ie MBHT) excels in modeling multi-behavior dependencies and outperforms all traditional baselines.
Regarding text-enhanced methods, UniSRec and VQ-Rec achieve better performance, thanks to its contrastive learning pre-training, which leads to high-quality universal sequence representation.
As for FDSA and S$^3$-Rec, they do not benefit from the text embeddings. 
One possible reason is the poor quality of text present on our platform, which requires the adoption of a robust text semantic modeling approach.
Finally, LSVCR consistently maintains the best results compared to the baseline methods.

\begin{table}[]
\centering
\caption{Performance comparison of various models on Amazon dataset. ``*'' denotes that the improvements are significant at the level of 0.01 with paired $t$-test.}
\label{tab:item_res}
\huge
\resizebox{1.0\linewidth}{!}{
\renewcommand\arraystretch{1.05}
\begin{tabular}{l|ccccc}
\hline
Methods & Recall@5 & Recall@10 & NDCG@5 & NDCG@10 & MRR \\
\hline
Caser    & 0.4277    & 0.5554     & 0.3178  & 0.3590   & 0.2985  \\
GRU4Rec  & 0.5143    & 0.6368     & 0.3968  & 0.4364   & 0.3743  \\
BERT4Rec & 0.4440    & 0.5665     & 0.3393  & 0.3789   & 0.3210  \\
SASRec   & 0.5111    & 0.6276     & 0.4011  & 0.4387   & 0.3802  \\
NARM     & 0.5121    & 0.6307     & 0.3983  & 0.4367   & 0.3765  \\
\hline
FDSA     & 0.5261    & 0.6422     & 0.4134  & 0.4509   & 0.3916  \\
S3Rec    & 0.5458    & 0.6631     & 0.4289  & 0.4669   & 0.4060  \\
UniSRec$_{t+ID}$  & \underline{0.5664}    & \underline{0.6822}     & \underline{0.4480}  & \underline{0.4855}   & \underline{0.4243}  \\
VQ-Rec  & 0.5324                       & 0.6567                        & 0.4267                     & 0.4601                     & 0.3999                  \\
\hline
\makecell[l]{ LSVCR\\[-0.5em] \large \ \ w/o Comment} & \textbf{0.6002*}    & \textbf{0.7088*}     & \textbf{0.4814*}  & \textbf{0.5166*}   & \textbf{0.4565*}  \\
\hline
\end{tabular}}
\end{table}

\noindent \textbf{Comment Recommendation.}
The results of the comment recommendation task are shown in Table~\ref{tab:res_com}. From these results, we can find:
DSSM, which does not involve user comment interaction history, exhibits inferior performance in comparison to other methods due to the lack of personalized user information. 
The personalized search methods that apply the title of the current video as a query (\ie ZAM, TEM) demonstrate better performance than the simple sequential modeling methods (\ie GRU, ATT, MHA, UniSRec$_{C}$).
Based on the findings, it is evident that the personalized comment recommendation task requires user interests closely aligned with the specific video content, rather than overall generalized user comment preference.
By comparing our approach with all baselines, it is clear that LSVCR achieves the most effective results. 


Overall, our LSVCR shows significant improvements over all baselines on both video and comment recommendation tasks.
This phenomenon indicates that joint modeling of videos and comments can enhance user video and comment preference learning simultaneously.
Through personalized preference alignment, we effectively integrate the powerful semantic understanding and knowledge reasoning capabilities of the LLM recommender into the SR model.
Furthermore, these results also show that the two-stage training paradigm can coordinate well for improving recommendation.

\subsection{Generalizability in Item Recommendation}
\label{sec:item_res}

In this part, we investigate whether the proposed personalized preference alignment is generalizable for other recommender systems.
However, to the best of our knowledge, there are no other public datasets consisting of comment data and interactions. 
Therefore, we employ the ``Arts, Crafts and Sewing'' subset of Amazon review data~\cite{ni2019justifying}, a general benchmark containing only item titles and user-item interactions, to validate our approach in item recommendation.
After processing, the dataset possesses about 20K items and 390K interactions. 
In this case, the techniques involving comment information in LSVCR will be discarded, including fusion encoders (Section~\ref{sec:seq_encode}), comment diversification (Section~\ref{sec:comdiv}), and video-comment preference contrast (Section~\ref{sec:vcc}). 
We name this variant of our proposed approach with LSVCR \underline{w/o Comment}.
Additionally, multi-behavior models (\ie DMT, MBHT) are also not applicable due to the absence of comment interaction behaviors.

The results are shown in Table~\ref{tab:item_res}. It can be seen that LSVCR outperforms all baseline models even after discarding the key techniques mentioned above. This indicates that the proposed two-stage training paradigm can effectively integrate the merits of the SR model and the supplemental LLM recommender. Besides, this phenomenon fully demonstrates that our LSVCR can be generalized to more general item recommendation scenarios.

\begin{table}[]
\centering
\caption{Ablation study on two recommendation tasks.
}
\label{tab:ablation}
\huge
\resizebox{0.98\linewidth}{!}{
\begin{tabular}{l|cc|cc}
\hline
\multirow{2}{*}{Methods} & \multicolumn{2}{c|}{Video Rec.}    & \multicolumn{2}{c}{Comment Rec.}           \\
\cline{2-5}
           & Recall@10  & NDCG@10 & Recall@10 & NDCG@10 \\
\hline
LSVCR & \textbf{0.3322} & \textbf{0.2233} & \textbf{0.3901} & \textbf{0.2541} \\
\ w/o $\mathcal{L}_{SSC}$  & 0.3211  & 0.2120 & 0.3822  & 0.2487 \\
\ w/o $\mathcal{L}_{VCC}$  & 0.3301  & 0.2211  & 0.3892   & 0.2528    \\
\ w/o ComDiv & 0.3280  & 0.2189    & 0.3853      & 0.2525  \\
\ w/o RandPosi   & 0.3267  & 0.2178 & 0.3881      & 0.2533   \\
\ w/o Alignment   & 0.3071 & 0.1998  & 0.3505   & 0.2176 \\
\ w/o $\mathcal{L}_{REG}$  & 0.3258   & 0.2171  & 0.3733  & 0.2428 \\
\ w/o Comment  & 0.3184 & 0.2097 & - & - \\
\ w/ BERT   & 0.3274 & 0.2181 & 0.3851 & 0.2511 \\
\ w/o $L_{LM}$  & 0.3206 & 0.2116 & 0.3759 & 0.2487 \\
\hline
\end{tabular}}
\end{table}

\subsection{Ablation Study}
In order to investigate how the proposed techniques affect the final performance, we conduct an ablation study on video and comment recommendation tasks. Specifically, we consider the following variants of LSVCR:
(1) {\underline{w/o $\mathcal{L}_{SSC}$}} without the sequential-supplemental preference contrast (Eq.~\eqref{eq:sscloss}).
(2) {\underline{w/o $\mathcal{L}_{VCC}$}} without the video-comment preference contrast (Eq.~\eqref{eq:vccloss}).
(3) {\underline{w/o ComDiv}} without the comment diversification augmentation (Section~\ref{sec:comdiv}) and ensuring inputs consistency across the two components.
(4) {\underline{w/o RandPosi}} without the randomized positional encoding (Section~\ref{sec:randposi}). It expands the positional encoding during fine-tuning and randomly initializes the expanded part.
(5) {\underline{w/o Alignment}} without the personalized preference alignment and learning the SR model directly.
(6) {\underline{w/o $\mathcal{L}_{REG}$ }} without the ID-text regularization loss (Eq.~\eqref{eq:regloss}).
(7) {\underline{LSVCR w/o Comment}} excludes comment information and simply aligns the SR model with the LLM recommender for video recommendation task (same as the variant in Section~\ref{sec:item_res}).
(8) {\underline{LSVCR w/ BERT}} utilizes a frozen BERT as text encoder and only trains LLM recommender in the alignment stage.
(9) {\underline{LSVCR w/o $L_{LLM}$}} without the target generation loss (Eq.~\eqref{eq:all_loss}).

The results presented in Table~\ref{tab:ablation} clearly demonstrate that removing any of the above techniques would lead to a decline in the overall effect.
Concretely, the absence of personalized preference alignment (\ie \underline{w/o Alignment}) results in significantly weaker performance.
Both contrastive losses (\ie $\mathcal{L}_{SSC}$ and $\mathcal{L}_{VCC}$) contribute to the downstream tasks.
Moreover, the performance degradation due to the absence of comment data once again demonstrates that comments can complement the video content, and comment interaction behaviors imply the user's interest.
It is noteworthy that our LSVCR achieves remarkable performance on both tasks, even when utilizing a frozen BERT encoder.
These observations highlight the importance of our proposed various techniques in integrating the merits of the SR model and the supplemental LLM recommender.

\begin{figure}[t]
\centering
\includegraphics[width=1.0\linewidth]{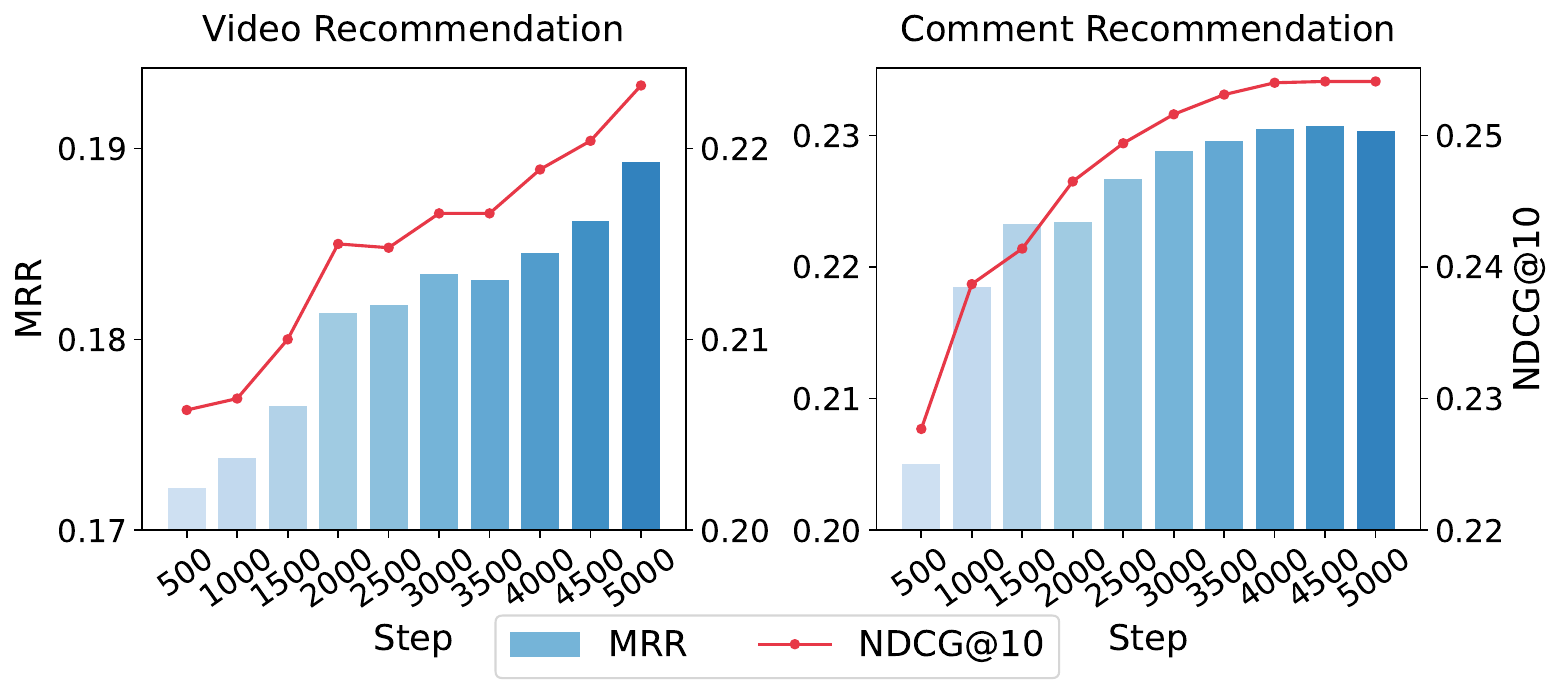}
\caption{Performance over different alignment steps.}
\label{fig:steps}
\end{figure}

\subsection{Further Analysis}

\subsubsection{Performance Comparison \wrt Alignment Steps}
In this section, we investigate how the number of personalized preference alignment steps affects performance. 
In the experiment, we utilize alignment-enhanced SR models with different numbers of alignment steps for fine-tuning.
From the results in Figure~\ref{fig:steps}, we can see that LSVCR primarily benefits from about the first 3000 alignment steps. 
For the video recommendation task, there is still considerable improvement in performance with continued training, which may be attributed to the availability of more diverse video recommendation data.
For the comment recommendation task, a duration of 3000 steps for alignment is deemed sufficient, which adequately showcases the superiority of LLM in swiftly adapting to new tasks.

\begin{figure}[]
\centering
\includegraphics[width=0.93\linewidth]{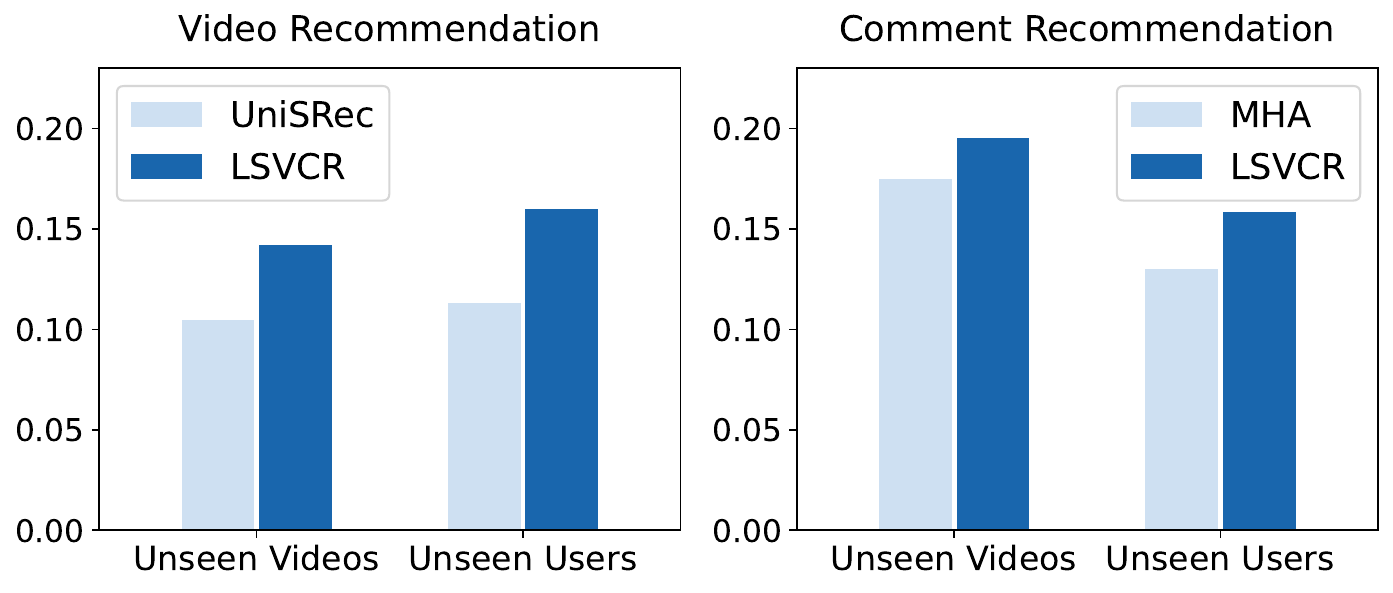}
\caption{NDCG@10 results of unseen videos and users}
\label{fig:unseen}
\end{figure}

\subsubsection{Performance Comparison \wrt Unseen Videos/Users.}

In this part, we attempt to apply our approach to videos or users that were not previously seen during training. 
Specifically, we first discard the ID embedding involved in our approach and baselines (\ie UniSRec, MHA), allowing adaptation in the presence of unseen videos. 
Then, we introduce two special datasets: (1) Unseen Videos: Select instances from the original test set where the target video never appeared during training. (2) Unseen Users: Select additional users who have never been seen and organize their interaction logs into a new dataset.
As shown in Figure~\ref{fig:unseen}, we can observe that LSVCR achieves more outstanding performance compared to the baselines on more challenging datasets in both video and comment recommendation tasks.
These findings indicate that with the assistance of personalized preference alignment between the SR model and the supplemental LLM recommender, LSVCR demonstrates robust capability in user preference modeling.

\subsubsection{Online A/B Testing}
In order to further verify the effectiveness of LSVCR, we deploy it to \emph{KuaiShou} platform for 14 days of online A/B testing.
Specifically, we incorporate our method into the re-ranking stage within the current recommendation workflow, and the test users are randomly divided into two groups for comparison. The experimental results are shown in Table~\ref{tab:abtest}.
We consider two metrics to measure user engagement:
(1) \emph{Watch Time}: Average video/comment watch time. (2) \emph{Interaction Num.}: Average number of video/comment interactions. Video interactions include ``like'', ``collect'' and ``share''. Comment interactions include ``like'' and ``reply''.
The results indicate that LSVCR achieves significant improvements on both video and comment recommendation tasks. 
It is worth noting that the promotion of comment recommendation is much more effective compared to video recommendation which may be attributed to the integration of LLM's powerful understanding capability on natural language in the proposed LSVCR.
Overall, our framework reveals the great potential of LLM for joint video and comment recommendation on online video platforms.

\begin{table}[]
\centering
\caption{Results of online A/B testing on \emph{KuaiShou} platform.}
\label{tab:abtest}
\renewcommand\arraystretch{1.03}
\begin{tabular}{l|cc}
\hline
Online Metrics   & Video & Comment  \\
\hline
Watch Time & +0.3649\%   & +4.1264\%  \\ 
Interaction Num.  & +0.7821\%     & +1.3557\% \\
\hline
\end{tabular}
\end{table}

\section{Conclusion}
In this paper, we proposed LSVCR that leverages user interaction histories with both videos and comments for joint video and comment recommendation.
There are two key components, namely conventional sequential recommendation~(SR) model and supplemental LLM recommender.
The former serves as the recommendation backbone, while the latter is used to enhance the semantics of the SR model.
A two-stage training paradigm is introduced to integrate the merits of both components.
In the first stage, we jointly learn the video and comment recommendation tasks while performing preference alignment between two components.
For the fine-tuning stage, we discard the supplemental LLM recommender and further fine-tune the alignment-enhanced SR model to improve performance on video and comment recommendation tasks.
Offline evaluation and online A/B testing on a real-world platform demonstrated the effectiveness of LSVCR.
As future work, we will explore more efficient ways to utilize LLM and attempt to incorporate a wider range of video and comment interaction behaviors in our framework.
Additionally, inspired by the potential of multi-modal large language models (MLLMs) in cross-modal modeling, we also intend to extend LLMs to MLLMs to integrate visual features.

\begin{acks}
This work was partially supported by National Natural Science Foundation of China under Grant No. 92470205 and 62222215, Beijing Municipal Science and Technology Project under Grant No. Z231100010323009, and Beijing Natural Science Foundation under Grant No. L233008. Xin Zhao is the corresponding author.
\end{acks}

\bibliographystyle{ACM-Reference-Format}

\balance
\bibliography{ref}

\end{document}